\documentclass[groupedaddress,prl,twocolumn,showpacs,amsmath,amssymb, amsfonts,10pt,superscriptaddress,noeprint]{revtex4-2}

\usepackage{amsmath}
\usepackage{amsfonts}
\usepackage{amssymb}
\usepackage{bbold}
\usepackage{mwe}
\usepackage{graphicx}
\usepackage{color}
\usepackage{bm}
\usepackage{soul}
\usepackage{braket}
\usepackage{xcolor}
\usepackage{placeins}
\usepackage[colorlinks,bookmarks=false,urlcolor=blue, citecolor=blue, linkcolor = blue]{hyperref}

\usepackage{comment}

\usepackage[capitalize]{cleveref}

\renewcommand{\vec}[1]{\boldsymbol{#1}}	
\renewcommand{\t}[1]{\text{#1}}

\newcommand{\varpm}{\mathbin{\vcenter{\hbox{\oalign{\hfil$\scriptscriptstyle+$\hfil\cr\noalign{\kern-.3ex}$\scriptscriptstyle({-})$\cr}}}}}
\newcommand{\varmp}{\mathbin{\vcenter{\hbox{\oalign{$\scriptscriptstyle({+})$\cr\noalign{\kern-.3ex}\hfil$\scriptscriptstyle-$\hfil\cr}}}}}

\newcommand{\dd}{\mathrm{d}}
\newcommand{\ee}{\mathrm{e}}

\definecolor{darkgreen}{rgb}{0.53, 0.66, 0.42}

\newcommand{\Hh}{\hat{H}}

\usepackage[normalem]{ulem}

\begin{document}

\title{Feshbach Resonances in Exciton--Charge-Carrier Scattering in Semiconductor Bilayers}

	\author{Marcel Wagner}
		\affiliation{Institute for Theoretical Physics, Heidelberg University, D-69120 Heidelberg, Germany}
    \affiliation{STRUCTURES, Heidelberg University, D-69117 Heidelberg, Germany}
	\author{Rafa{\l} O{\l}dziejewski}
	\affiliation{Max-Planck-Institute of Quantum Optics, D-85748 Garching, Germany}
	\author{F\'elix Rose}
	\affiliation{Institute for Theoretical Physics, Heidelberg University, D-69120 Heidelberg, Germany}
    \affiliation{STRUCTURES, Heidelberg University, D-69117 Heidelberg, Germany}
    \affiliation{Laboratoire de Physique Th\'eorique et Mod\'elisation, CY Cergy Paris Universit\'e, CNRS, F-95302 Cergy-Pontoise, France}
    \author{Verena K\"{o}der}
	\affiliation{Institute for Theoretical Physics, Heidelberg University, D-69120 Heidelberg, Germany}
	\author{Clemens Kuhlenkamp}
	\affiliation{Institute for Quantum Electronics, ETH Z\"urich, CH-8093 Z\"urich, Switzerland}
    \affiliation{Department of Physics, Technical University of Munich, D-85748 Garching, Germany}
	\author{Ata\c{c} \.{I}mamo\u{g}lu}
	\affiliation{Institute for Quantum Electronics, ETH Z\"urich, CH-8093 Z\"urich, Switzerland}
	\author{Richard Schmidt}
	\affiliation{Institute for Theoretical Physics, Heidelberg University, D-69120 Heidelberg, Germany}
	\affiliation{STRUCTURES, Heidelberg University, D-69117 Heidelberg, Germany}

	\date{\today}

	\begin{abstract}
		Feshbach resonances play a vital role in the success of cold atoms investigating strongly correlated physics. The recent observation of their solid-state analog in the scattering of holes and intralayer excitons in transition metal dichalcogenides [I. Schwartz \textit{et al.}, Science \textbf{374}, 336 (2021)] %~\cite{Schwartz2021} 
        holds compelling promise for bringing fully controllable interactions to the field of semiconductors. Here, we demonstrate how tunneling-induced layer hybridization can lead to the emergence of two distinct classes of Feshbach resonances in atomically thin semiconductors. Based on microscopic scattering theory we show that these two types of Feshbach resonances allow us to tune interactions between electrons and both short-lived intralayer, as well as long-lived interlayer excitons. We predict the exciton-electron scattering phase shift from first principles and show that the exciton-electron coupling is fully tunable from strong to vanishing interactions. The tunability of interactions opens the avenue to explore Bose-Fermi mixtures in solid-state systems in regimes that were previously only accessible in cold atom experiments. 
	\end{abstract}

	\maketitle

The past years have seen the advent of few-layer transition metal dichalcogenides (TMDs) as a novel platform to study strongly correlated quantum matter~\cite{sidler2017fermi,wang2018tmdspectroscopy}, marked by the observation of Mott phases~\cite{Regan2020,Shimazaki2021}, insulating density waves~\cite{Xu2020,Jin2021,Huang2021}, excitonic insulators~\cite{Amelio2022}, Wigner crystals~\cite{Regan2020,smolenski2021,zhou2021}, the quantum anomalous Hall effect \cite{Xie2022}, and fractional Chern insulators \cite{Cai2023}.  In TMDs, Bose-Fermi mixtures of excitons and electrons reach for the first time into the previously inaccessible strong-coupling regime \cite{Fey2020,sidler2017fermi,Cotlet2016,Efimkin218,Imamoglu2021}. While such mixtures have allowed for the optical detection and engineering of novel many-body phases \cite{smolenski2021}, the control over exciton-electron interactions has so far been limited to doping \cite{Katsch2022} or coupling to photonic modes in optical cavities \cite{Li2021,Li2021b,Kumar2023,Rana2021,Bastarrachea-Magnani2021} and wave guides \cite{Koksal2021}.
 
 Establishing fully tunable interactions of electrons with both short- and long-lived excitons would open up the possibility to explore even richer many-body phases such as exciton-induced superconductivity \cite{laussy2010} or charge-density wave states \cite{Shelykh2010,Cotlet2016}, and would further enrich TMDs as a fully tunable quantum simulation platform on par with ultracold atoms.

In this Letter, we show how full control over interactions can be realized in TMDs. Using a quantum-chemistry inspired microscopic approach we demonstrate that the presence of trions both in open and closed scattering channels allows for the full tunability of interactions from the %\cst{ultra-}
strong coupling regime all the way to vanishing interactions both for intra- and interlayer excitons. The trions in the respective closed channel take the role analogous to closed-channel molecules in cold atoms where they build the foundation of tunable interactions between atomic particles \cite{Chin2010}. In contrast to cold atoms, however, a clear scale separation between the energy of the closed-channel bound states and the relevant scattering energies is absent. This renders the three-body nature of the underlying scattering processes a crucial ingredient which cannot be captured by effective theories based on structureless particles~\cite{Kuhlenkamp2022}.

Here we address this challenge by starting from a microscopic model that fully resolves the internal structure of the three-particle complexes underlying Feshbach-enhanced exciton-electron scattering. Our solution of the quantum three-body problem is based on exact diagonalization and reveals the existence of two types of resonances. The first type of resonance has the characteristics of broad Feshbach resonances in ultracold atoms. This resonance allows to tune the interactions between electrons and short-lived intralayer excitons of large oscillator strength, making them ideal for applications in spectroscopy and correlation sensing. 

The second type of resonance couples long-lived interlayer excitons and electrons. It has the characteristics of a narrow resonance and requires fine-tuning of the external electric field representing the control parameter of the Feshbach resonances in TMD heterostructures. This novel type of resonance allows the realization of long-lived exciton-electron mixtures at strong-coupling, and thus enables a new approach to explore the phase-diagram of Bose-Fermi mixtures in regimes that have so far been out of reach in cold atomic systems due to their chemical instability \cite{Duda2023}. 
	
\begin{figure}[t!]
\centering
		\includegraphics{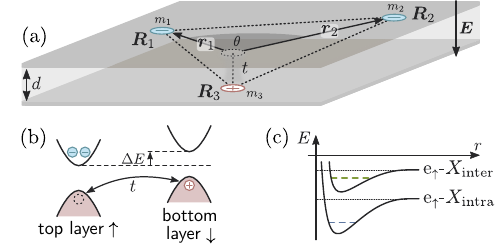}
\caption{(a) Sketch of the bilayer three-body system. The hole (red) can tunnel between the layers with a coupling constant $t$, whereas electrons (blue) are confined to the top layer. The lab frame coordinates of the charges are $\vec{R}_i$, $i=1,2,3$. The relative positions $\vec{r}_1$ and $\vec{r}_2$ span an angle $\theta$. (b) Illustration of the band structure detuning where the layer index takes the role of a pseudospin degree of freedom $\ket{\uparrow}$, $\ket{\downarrow}$. (c) Sketch of the effective exciton-electron scattering potentials, each supporting a trion state (dashed).}
\label{fig:model}
\end{figure}
	
	\textbf{\textit{Setup and model.---}} 
We consider a bilayer system where two electrons in one layer interact with a hole that can tunnel in between the layers~(cf.~\cref{fig:model}). The layers are separated by a single sheet of hexagonal boron nitride (hBN) of thickness 
 $0.33\,$nm \cite{Zhang2017} resulting in a spacing of $d=1.03\,$nm of the central planes of two MoSe${}_2$ layers of thickness $0.7\,$nm \cite{Borodin2018}.
The layers define a pseudospin degree of freedom: akin to a cold atom system, the interactions between particles depend on this pseudospin, which in the TMD setting are coupled through tunneling. The system is subject to an external perpendicular electric field $\mathbf{E} = E_z \hat{ \mathbf e}_z$. This field allows to tune the energy difference $\Delta E = e E_z d$ between the pseudospin states, analogous to magnetic fields employed to realize Feshbach resonances in cold atoms. 
In an effective mass approach the corresponding Hamiltonian reads
	\begin{equation}\label{eq:hamiltonian_abinitio}
	\Hh = \sum_i \bigg[\frac{\hat{\vec{p}}_i^2}{2 m_i} - \frac{Q_i \Delta E}{2} (\mathbb{1}-\tau_z^i) + t_i \tau_x^i\bigg]
 +\sum_{\overset{{}_{\phantom{A}} \!\!\!\! \scriptstyle{a,b,}}{\scriptstyle{i<j}}} V_{ab}(\hat{R}_{ij})
	\end{equation}
where $Q_i=\pm e$ and $m_i$ are, respectively, the charges and the masses of the electrons ($i=1,2$) and the hole ($i=3$). The in-plane distances and momenta of the particles are described by $\hat{R}_{ij}=|\hat{\vec{R}}_i-\hat{\vec{R}}_j|$ and $\hat{\vec{p}}_i$, respectively. The Pauli matrices $\tau_{\mu=x,y,z}^i$ act on the layer (pseudospin) subspace of each particle, labeled by $a,b\in \{\uparrow,\downarrow\}$, where the ``pseudospin" $\uparrow,\downarrow$ refers to the layer index in which the respective charge carrier resides.

The first term in \cref{eq:hamiltonian_abinitio} represents the kinetic energy of particles and the second the energy detuning $\Delta E$ of the two layers. The third term accounts for the tunneling of the hole, $t_3=t$ and $t_1=t_2=0$. This choice of species-dependent tunnel coupling is motivated by experiments \cite{Schwartz2021} and may be seen as originating from the energy offset between the conduction bands of the hBN and TMD layers, which is drastically larger than the one of the valence bands. The resulting confinement of each electron to one layer leads to the decoupling of Hilbert space into invariant subspaces.
In the following we focus on the case where the electrons are located in the top layer.

The interaction potential $V_{ab}(r)$ between two charges separated by an in-plane distance $r$ is obtained from solving Poisson's equation for two identical  TMD layers (of vanishing thickness) separated by the distance $d$. %that models the thickness of a monolayer of hBN. 
The interaction potential in momentum space reads 

 \begin{equation}\label{intrapot}
 V_{\text{intra}}(q)=\frac{Q_i Q_j\left((1+r_{0}q)\ee^{2dq}-r_{0}q\right) }{2\varepsilon_0\left((1+r_{0}q)^{2} q \ee^{2dq} - r_0^2 q^3\right)}
 \end{equation}
 for two charges in the same layer ($a=b$), and 
 \begin{equation}\label{interpot}
V_{\text{inter}}(q)=\frac{Q_i Q_j \ee^{dq}}{2\varepsilon_0\left((1+r_{0}q)^{2} q \ee^{2dq} - r_0^2 q^3\right) }
 \end{equation}
for charges in different layers ($a\neq b$; for details see the Supplemental Material \cite{SM}). The real-space potentials are obtained via Fourier transformation. The screening length is $r_0=\alpha_{2\text{D}}/2\varepsilon_0$ with the TMD layer's two-dimensional polarizability $\alpha_{2\text{D}}$ and the vacuum permittivity $\epsilon_0$. Equations \eqref{intrapot} and~\eqref{interpot} reduce to the Keldysh potential \cite{Keldysh1979} in the ultrathin, monolayer limit $d\to 0$ \cite{SM}.

Importantly, the three-body model in \cref{eq:hamiltonian_abinitio} does not \textit{a priori} assume the formation of excitons nor does it treat them as rigid objects. Instead, excitons appear, on equal footing to the trions, as eigenstates of the Hamiltonian and can themselves be layer hybridized due to charge-carrier tunneling. 
	
\textbf{\textit{Feshbach resonances.---}} 
To understand the exciton-electron scattering physics, we diagonalize the Hamiltonian \eqref{eq:hamiltonian_abinitio} using a discrete variable representation (DVR)~\cite{Beck2000,Fey2020}. Exploiting translational invariance, we transform the Hamiltonian into the co-moving frame of the hole and separate the center-of-mass motion \cite{SM}. Rotational invariance implies the conservation of total angular momentum $m$. In the following we focus on $m=0$. The eigenstates of the resulting Hamiltonian are wave functions $\psi(r_1,r_2,\theta)$ that, additionally to the layer degrees of freedom, are parametrized by relative particle distances $r_{1,2}$ and the angle $\theta$ between the electron coordinates, see \cref{fig:model}. While our approach describes generally electrons and holes in any TMD heterostructure, we assume two MoSe${}_2$ layers and use material parameters obtained from DFT~\cite{Kylanpaa2015}. We expect our results to be quite universal and to apply to other material combinations as well. 
 
	\begin{figure}
        \centering
		\includegraphics{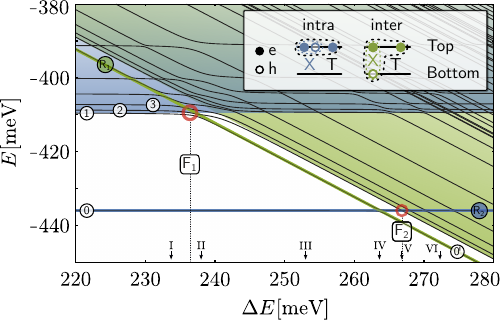}
		\caption{Spectrum of three-body states, spatially symmetric with respect to electron exchange, as function of the band detuning $\Delta E$. Bold colored lines mark trion energies for $t=0$ whereas black lines show the eigenenergies (including trion and exciton-electron scattering states) in presence of hole tunneling. 
 Shaded areas  mark the different scattering states which form a continuum in the infinite-system limit. Blue color indicates intralayer, and green color interlayer configurations as shown in the pictogram. Feshbach resonances appear at the positions labeled by F${}_1$ and F${}_2$. Symbols I-VI mark $\Delta E$ values for which the states labeled by 0-3, 0${}^{\prime}$, R${}_1$ and R${}_2$ are visualized in \cref{fig:States}. 
 }
		\label{fig:SymmSpectrumE1}
	\end{figure}

 The exact diagonalization gives access to a rich spectrum containing both symmetric and antisymmetric states with respect to electron exchange. In \cref{fig:SymmSpectrumE1} we show the result for the spatially symmetric states (i.e.\, unequal spin or valley degree of freedom) and study how the low-energy spectrum depends on the band detuning $\Delta E$. We mark the energies of trions in the absence of tunneling with bold colored lines. Depending on whether the hole is in the top or in the bottom layer two such trion states exist. We denote these as bare intralayer (blue) or bare interlayer trion (green). The shaded areas mark the corresponding electron-exciton scattering continua. We find that the hole is always part of the excitonic component of the wave function. Since increasing $\Delta E$ decreases the energy of the hole, one is able to bring the originally more weakly bound interlayer exciton into resonance with the intralayer exciton state. The thin black lines in Fig.~\ref{fig:SymmSpectrumE1} mark the energies of the eigenstates of the system in presence of a finite hopping strength, $t=2$ meV~\cite{Schwartz2021}. They can be expressed as a superposition of the eigenstates for $t=0$.  
% \cst{The states belonging to the scattering continua appear as sets of discrete states due to the finite size of our system.} 
In our numerics the system size is finite. As a result the   spectrum of scattering states that is continuous in the infinite system size limit (shaded) appears as a discrete set of states (lines) in the numerics.
 
 By following the energy of bare trions (`closed-channel') one can identify points where they cross the exciton-electron scattering threshold (`open-channel'). Tunneling of the hole couples these channels. This turns the closed-channel trions into weakly bound Feshbach trions, leading to the emergence of the Feshbach resonances marked by the labels F${}_1$ and F${}_2$.

	\begin{figure}
       \centering
		\includegraphics{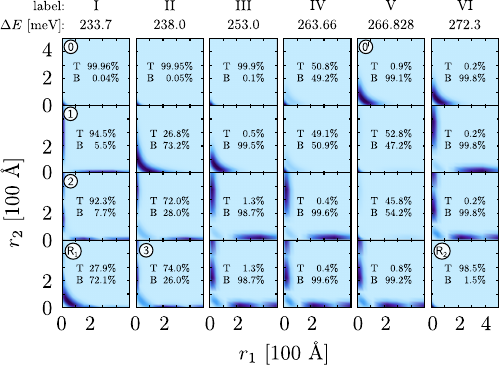}
		\caption{Angle- and layer-averaged probability densities of electrons $n(r_1,r_2)\propto\int \dd \theta\, |\psi(r_1,r_2,\theta)|^2$ in the lowest states at the positions marked with labels I-VI in \cref{fig:SymmSpectrumE1}. The layer hybridization of intra- (T: hole in top layer) and interlayer (B: hole in bottom layer) configurations is shown for each state. 
		}
		\label{fig:States}
	\end{figure}
 
In the emergence of exciton-electron Feshbach resonances, layer hybridization plays a key role. In \cref{fig:States} we show the layer- and $\theta$-averaged probability density $n(r_1,r_2)\propto\int \dd \theta\, |\psi(r_1,r_2,\theta)|^2$ of the lowest states for the band detunings $\Delta E$ labeled as I-VI in \cref{fig:SymmSpectrumE1}. 
For each state we show the degree of layer hybridization as probability of finding the hole in the top (T) or bottom (B) layer. The upper most row shows the lowest trion state. As $\Delta E$ is increased to a value around %\cst{$\Delta E\approx 140$} \new{
 $\Delta E\approx 264$ meV the deeply bound intralayer trion (label $0$ in Figs.~\ref{fig:SymmSpectrumE1} and \ref{fig:States}) turns into an interlayer trion (label $0'$). The second and following rows in Fig.~\ref{fig:States} show excited states (labels $1$-$3$). For %\cst{$\Delta E\approx 120$} \new{
 $\Delta E = 233.7$ meV (I) the multinodal structure of the wave function for one of the electrons shows that these states represent intralayer exciton-electron scattering states. 
 The lowest left subfigure in Fig.~\ref{fig:States} shows the interlayer trion (label $R_1$) that is immersed in the intralayer scattering continuum. This trion is subject to layer hybridization turning it into a metastable, resonant state. It becomes increasingly unstable as the broad Feshbach resonance F${}_1$ is approached.
 
When crossing the resonance F${}_1$ at %\cst{$\Delta E\approx 125$} \new{
$\Delta E \approx 236.4$ meV, we observe a transition of the first excited state (second row in \cref{fig:States}) from being the lowest intralayer scattering state into the newly emergent Feshbach trion. 
Only at larger values of $\Delta E$ this intralayer-dominated trion then turns into the interlayer trion as highlighted by the complete reorganization of layer hybridization in this state. At the same time, the second intralayer scattering state (showing a binodal structure) turns into the first intralayer scattering state (showing only one node) while the ground state wave function remains nearly unchanged (upper row). Analogously, higher excited states change their number of radial nodes by one when crossing the resonant state. 

A further increase of $\Delta E$ beyond F${}_1$ leads to the anticrossing of the trions at %\cst{$\Delta E\approx 139$} \new{
$\Delta E\approx 263.66$ meV (label IV), accompanied by a maximal layer hybridization. Because of the bound character of the states this hybridization is robust and independent of the system size of our numerical diagonalization. The intralayer trion crosses the interlayer scattering continuum at the second resonance $\text{F}_2$ at %\cst{$\Delta E\approx 141.7$} \new{
$\Delta E \approx 266.828$ meV (label V). At this point it turns into a hybridized, resonant state before recovering its pure intralayer character only for larger $\Delta E$ detunings (state $R_2$ in Figs.~\ref{fig:SymmSpectrumE1} and \ref{fig:States}). We note that the layer hybridization found in Fig.~\ref{fig:States} implies a modification of the electric dipole of the excitons and trions. This might give rise to interesting, tunable many-body physics of excitons and trions at finite density. 

We now turn to the detailed analysis of the Feshbach resonances by investigating the scattering properties of the lowest scattering state (open channel) when tuning $\Delta E$ across the resonances. Specifically, we determine the phase shift of the electron-exciton scattering which determines the strength of the modification of the scattering states due to interactions. To this end we fit the long-distance part of the lowest scattering state to the asymptotic scattering wave function, $R_n(r)=\alpha_m(k_n)J_m(k_n r)+\beta_{m}(k_n)Y_m(k_n r)$. Here $J_m$ and $Y_m$ are the Bessel functions of the first and second kind, respectively, and $k_n$ the momentum of the $n$th scattering state. We obtain the 2D scattering phase shift from $\delta_{m}(k_n)=-\t{arctan}\left(\beta_{m}(k_n)/\alpha_{m}(k_n) \right)$ ~\cite{Adhikari1986,Verhaar1984}.

The low-energy $s$-wave scattering phase shift $\delta_E \equiv \delta_0(k)$ (with $E = \hbar^2 k^2/\mu$) is parameterized in terms of the energy scale $E_{\t{s}}$ according to $\t{cot}(\delta) \overset{E\rightarrow 0}{=} 1/\pi\, \t{ln} ( E/E_{\t{s}} )$. The scale $E_{\t{s}}$ can be linked to the 2D scattering length $a_{\t{2D}}$, conventionally used in the context of cold atom experiments, via $a_{\t{2D}} = \sqrt{\hbar^2/2 E_{\t{s}} \mu}$. It can be interpreted as the characteristic kinetic energy scale on which a many-body system experiences resonant, strong coupling physics. For instance, in fermionic systems the Fermi energy $E_F$ would provide such a scale.

 \begin{figure}[t!]
		\centering
  \includegraphics[width=1\linewidth]{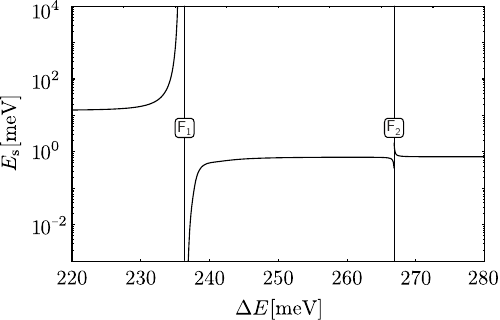}
		\caption{Band detuning dependence of the exciton-electron scattering parameter $E_{\t{s}}$. The broad and narrow Feshbach resonances are labeled as $\text{F}_1$ and $\text{F}_2$, respectively (as also shown in \cref{fig:SymmSpectrumE1}). }
		\label{fig:Es}
	\end{figure}
 
We show the scattering parameter $E_{\t{s}}$ in Fig.~\ref{fig:Es}. 
Far away from the resonance $\text{F}_1$, $E_{\t{s}}$ is approximately constant and takes a value that is on the order of the respective background trion energy (intralayer on the left and interlayer on the right of $\text{F}_1$), with deviations due to the details of the short-distance scattering physics.
Approaching $\text{F}_1$ from the left $E_{\t{s}}$ increases and diverges. Correspondingly, at low scattering energies the system becomes effectively noninteracting. Right on resonance, $E_{\t{s}}$ jumps abruptly to zero and a bound state appears. 
In this regime $E_{\t{s}}$ follows precisely the energy of the newly emerging Feshbach bound trion. Note, on the left-hand side of $\text{F}_1$, we evaluate $E_s$  with respect to the  electron-intralayer exciton scattering channel, while, on the right-hand side of $\text{F}_1$, $E_s$ is evaluated with respect to the electron-interlayer exciton scattering channel. This switch of the branch of the scattering channel leads to the ``gap" in values  for $E_s$ seen in Fig.~\ref{fig:Es}.

The equivalence of vanishing and diverging scattering lengths at F${}_1$ and F${}_2$ is a hallmark of 2D scattering physics \cite{Kanjilal2006}. In contrast to 3D Feshbach resonances in ultracold atoms, thus also the concept of resonance width, defined by the distance in magnetic field of the zero crossing of the scattering length to the resonance position \cite{Chin2010}, is not applicable in TMDs. Namely, the zero of $a_{\t{2D}}$ actually coincides precisely with the 2D Feshbach resonance position. This scattering regime, that is a unique 2D feature and has remained inaccessible in cold atoms even when using confinement-induced resonances~\cite{Chin2010}, can now be realized and studied in atomically thin semiconductors.

The hybridization of intra- and interlayer scattering thresholds in the vicinity of $\text{F}_1$ facilitates a large coupling between the open and closed scattering channels. This results in the broad Feshbach resonance at $\text{F}_1$. In contrast, the resonance at $\text{F}_2$ is much more narrow in terms of the $\Delta E$ change required to tune the system across the resonance (if the tunnel coupling is increased, the modification would be stronger). Importantly, the character of the resonances $\text{F}_1$ and $\text{F}_2$ differs with respect to the lifetime and oscillator strength of excitons in the respective open channel. While $\text{F}_1$ allows to manipulate the scattering of electrons and short-lived intralayer excitons with a large oscillator strength, the scattering physics of electrons with long-lived intralayer excitons (of small oscillator strength) can be tuned via $\text{F}_2$.
This allows us to use intralayer excitons, optically injected close to $\text{F}_1$, as a tunable probe for correlation sensing of electronic many-body systems, while the resonance at F${}_2$ brings tunable interactions to many-body systems comprised of electrons and stable interlayer excitons, paving the way to controllable, long-lived 2D Bose-Fermi mixtures. 

Note, in this work we have focused on one specific configuration that is directly relevant for ongoing experiments. Allowing for different configurations (including modified tunneling strength, charge, valley, and spin degrees of freedom) will give rise to an even richer set of Feshbach resonances to be explored. 

\textbf{\textit{Conclusion.---}} In this work we have studied the emergence of electrically tunable 2D Feshbach resonances that allow to tune the scattering properties of short- and long-lived excitons with electrons. In both cases the systems can be tuned to the limit of vanishing interactions. Consequently, these 2D Feshbach resonances might pave the way to probe many-body physics using valley-selective interferometric protocols, such as Ramsey or spin-echo schemes, applied to excitons~\cite{cetina2016ultrafast}. Our findings may open up the prospect to realize many-body systems comprised of long-lived, dipolar interlayer excitons and itinerant electrons, where tunable interactions might enable exciton-induced superconductivity \cite{laussy2010,Laussy2012} or supersolidity in dipolar exciton condensates \cite{Shelykh2010,Cherotchenko2016,Cotlet2016}.

\textit{Acknowledgements.---}
We thank Arthur Christianen, Christian Fey, Jesper Levinsen, Meera Parish and Jonas von Milczewski for inspiring discussions. M.\,W., F.\,R., and R.\,S. acknowledge support by the Deutsche Forschungsgemeinschaft under Germany's Excellence Strategy EXC 2181/1 - 390900948 (the Heidelberg STRUCTURES Excellence Cluster), and CRC 1225 ISOQUANT, project-ID 273811115. M.\,W. and R.\,S. acknowledge also support by DFG Priority Programme of Giant Interactions in Rydberg Systems (GiRyd). The work at ETH Zurich was supported by the Swiss National Science Foundation (SNSF) under Grant Number 200021-204076.

\end{document}

\title{Supplemental Material:\ Feshbach Resonances in Exciton--Charge-Carrier Scattering in Semiconductor Bilayers}

	\author{Marcel Wagner}
		\affiliation{Institute for Theoretical Physics, Heidelberg University, D-69120 Heidelberg, Germany}
    \affiliation{STRUCTURES, Heidelberg University, D-69117 Heidelberg, Germany}
	\author{Rafa{\l} O{\l}dziejewski}
	\affiliation{Max-Planck-Institute of Quantum Optics, D-85748 Garching, Germany}
	\author{F\'elix Rose}
	\affiliation{Institute for Theoretical Physics, Heidelberg University, D-69120 Heidelberg, Germany}
    \affiliation{STRUCTURES, Heidelberg University, D-69117 Heidelberg, Germany}
    \affiliation{Laboratoire de Physique Th\'eorique et Mod\'elisation, CY Cergy Paris Universit\'e, CNRS, F-95302 Cergy-Pontoise, France}
    \author{Verena K\"{o}der}
	\affiliation{Institute for Theoretical Physics, Heidelberg University, D-69120 Heidelberg, Germany}
	\author{Clemens Kuhlenkamp}
	\affiliation{Institute for Quantum Electronics, ETH Z\"urich, CH-8093 Z\"urich, Switzerland}
    \affiliation{Department of Physics, Technical University of Munich, D-85748 Garching, Germany}
	\author{Ata\c{c} \.{I}mamo\u{g}lu}
	\affiliation{Institute for Quantum Electronics, ETH Z\"urich, CH-8093 Z\"urich, Switzerland}
	\author{Richard Schmidt}
	\affiliation{Institute for Theoretical Physics, Heidelberg University, D-69120 Heidelberg, Germany}
	\affiliation{STRUCTURES, Heidelberg University, D-69117 Heidelberg, Germany}

	\date{\today}
	
	\graphicspath{{./Figures/}}

\maketitle

\setcounter{equation}{0}
\setcounter{figure}{0}
\setcounter{table}{0}
\setcounter{page}{1}
\makeatletter
\renewcommand{\theequation}{S\arabic{equation}}
\renewcommand{\thefigure}{S\arabic{figure}}
\renewcommand{\bibnumfmt}[1]{[#1]}
\renewcommand{\citenumfont}[1]{#1}

\section{Electrostatic interactions of charge carriers in two-layer systems of two-dimensional materials}
\label{SM:Potential}
In this section, we derive the electrostatic potentials used in the main text of this work. We therefore extend the approach in Ref.~\cite{Cudazzo2011} to a two layers of two-dimensional materials in the ultrathin limit that are separated by the distance of hexagonal boron nitride (hBN). Note, that we obtain the very same interaction potentials applying the boundary conditions of electrostatic fields on surfaces on finitely thick sheets, taking the two-dimensional limit
in the end. In order to simplify the analysis, we approximate the hBN layer to be dielectrically isotropic with $\varepsilon_{\perp}=\varepsilon_{\parallel}=\varepsilon_0$, the vacuum permittivity, but keeping the respective layer separation. We expect this approximation to not influence our result qualitatively.

\begin{figure}[]
		\centering
		\includegraphics[width=0.36\textwidth]{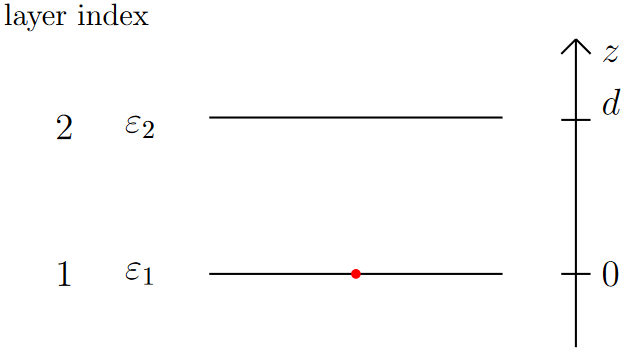}
		\caption{Sketch of the scenario considered in the calculation of the electrostatic potential. The two parallel layers with indices 1 and 2 are of zero thickness and are separated by a distance $d$ in $z$-direction. They carry the permittivity $\varepsilon_1$ and $\varepsilon_2$, respectively and are embedded in vacuum with permittivity $\varepsilon_0$. We consider the potential caused by a point charge $Q$ in layer 1, as it is felt by a test charge in either layer 1 or 2. }
		\label{fig:ElectrostatPotential}
\end{figure}

The scenario considered is depicted in \cref{fig:ElectrostatPotential}. To begin with, we consider a point-like particle with charge $Q$ positioned at the origin of the coordinate system. In the following, we will describe the test charge's position by $\Vec{x}=(\Vec{r},z)^T$ and also label the directional derivatives accordingly. Using the relations $\rho_{\text{ind}}=-\nabla \Vec{P}_{2D}, \, \Vec{P}_{2D}=-\varepsilon_0 \alpha_{2D}\nabla_{\Vec{r}}\Phi(\Vec{x})\delta(z)$ for the induced charge density and the polarization of a two-dimensional layer, where $\alpha_{2D}$ is the two-dimensional polarizability, we can write Poisson's equation as
\begin{equation}
\Delta \Phi(\Vec{x})=-\frac{1}{\varepsilon_0}\left(\rho_{\text{ext}}+\rho_{\text{ind}_1}+\rho_{\text{ind}_2}\right)=-\frac{1}{\varepsilon_0}\big(\delta^3(\Vec{x}) Q+ \delta(z) \varepsilon_0\alpha_{2D_1}\Delta_r \Phi(\Vec{x})+\delta(z-d)\varepsilon_0\alpha_{2D_2}\Delta_r \Phi(\Vec{x})\big) \,.
\end{equation}
After Fourier transformation in both $\Vec{r}$ and $z$, this can be rewritten as
\begin{equation}\label{eq:potFT}
\begin{aligned}
    &\Phi(\Vec{q},q_z)=\frac{1}{\varepsilon_0(|\Vec{q}|^2+q_z^2)} \bigg(Q-\varepsilon_0\alpha_{2D_1} |\Vec{q}|^2 \Phi_{2D}(\Vec{q},z=0)-\varepsilon_0\alpha_{2D_2} |\Vec{q}|^2 \Phi_{2D}(\Vec{q},z=d) \ee^{-\ii q_zd}\bigg)\,,\\
    &\text{with} \quad \Phi_{2D}(\Vec{q},z=a)=\int \frac{\mathrm{d}q_z}{2 \pi} \Phi(\Vec{q},q_z) \ee^{\ii q_za}\,.
\end{aligned}
\end{equation}
Integrating over $q_z$, in order to express \cref{eq:potFT} in terms of $\Phi_{2D}$, allows us to investigate the potential at $z=0$ and $z=d$, which are the two possible z-positions of the test charge. From now on we use $p\equiv|\Vec{p}|$. For $z=0$, we obtain
\begin{equation}
\begin{aligned}
\Phi_{2D}(q,0)&=\int_{-\infty}^{\infty} \frac{\mathrm{d}q_z}{2\pi \varepsilon_0}\frac{Q-\varepsilon_0 \alpha_{2D_1}q^2\Phi_{2D}(q,0)-\varepsilon_0\alpha_{2D_2}q^2\Phi_{2D}(q,d)\ee^{-\ii q_zd}}{q^2+q_z^2} \\
&=\int_{-\infty}^{\infty}\frac{\mathrm{d}\xi}{2\pi \varepsilon_0} \frac{Q/q-\varepsilon_0\alpha_{2D_1} \Phi_{2D}(q,0) q-\varepsilon_0\alpha_{2D_2} \Phi_{2D}(q,d) q \ee^{-\ii q\xi d}}{1+\xi^2}\\
&=\frac{1}{2\pi \varepsilon_0}\bigg[\frac{Q}{q}\left[ \mathrm{arctan}(\xi)\right]_{-\infty}^{\infty}- \varepsilon_0\alpha_{2D_1} q \Phi_{2D}(q,0)\left[ \mathrm{arctan}(\xi)\right]_{-\infty}^{\infty}+ 2 \pi i \, \mathrm{Res}_{\,\Xi=-\ii \!}\left( \frac{ q \varepsilon_0\alpha_{2D_2} \Phi_{2D}(q,d)}{1+{\Xi}^2} \ee^{-\ii q\Xi d}\right)\bigg] \\
&=\frac{1}{2\varepsilon_0}\left(\frac{Q}{q}-\varepsilon_0\alpha_{2D_1}q\Phi_{2D}(q,0)-\varepsilon_0\alpha_{2D_2} q\Phi_{2D}(q,d) \ee^{-dq}\right)\,,
\end{aligned}
\end{equation}
where, in the second line, we defined $\xi\equiv q_z/q$, and $\Xi$ denotes $\xi$'s extension to the complex plane. 
For $z=d$, we get
\begin{equation}
\begin{aligned}
\Phi_{2D}(q,d)&=\int_{-\infty}^{\infty}\frac{ \mathrm{d}q_z \ee^{\ii q_z d}}{2\pi \varepsilon_0}\frac{Q-\varepsilon_0\alpha_{2D_1}q^2\Phi_{2D}(q,0)-\varepsilon_0\alpha_{2D_2}q^2\Phi_{2D}(q,d)\ee^{-\ii q_zd}}{q^2+q_z^2} \\
&=\int_{-\infty}^{\infty} \frac{\mathrm{d}\xi}{2\pi \varepsilon_0} \frac{\ee^{\ii q\xi d}(Q/q-\varepsilon_0\alpha_{2D_1} q\Phi_{2D}(q,0) )-\varepsilon_0\alpha_{2D_2} q\Phi_{2D}(q,d) }{1+\xi^2} \\
&=\frac{1}{2\pi \varepsilon_0}\bigg(2 \pi i \, \bigg[\mathrm{Res}_{\,\Xi=\ii \!}\left( \frac{Q}{q}\frac{\ee^{\ii q \Xi d}}{1+{\Xi}^2}\right)-\varepsilon_0 \alpha_{2D_1} \, \mathrm{Res}_{\,\Xi=\ii \!}\left( \frac{q\Phi(q,0)}{1+{\Xi}^2}\ee^{\ii q\Xi d}\right)\bigg]-\varepsilon_0\alpha_{2D_2}q\left[ \mathrm{arctan}(\xi)\right]_{-\infty}^{\infty}\Phi(q,d)\bigg) \\
&=\frac{1}{2\varepsilon_0}\left(\frac{Q}{q}\ee^{-qd}-\varepsilon_0\alpha_{2D_1}q \ee^{-qd} \Phi_{2D}(q,0)-\varepsilon_0\alpha_{2D_2}q\Phi_{2D}(q,d)\right) \, ,
\end{aligned}
\end{equation}
which leads to
\begin{equation} \label{eq:2layer1}
\Phi_{2D}(q,0)=\frac{Q/q-\varepsilon_0\alpha_{2D_2}q \ee^{-dq}\Phi_{2D}(q,d) }{2 \varepsilon_0(1+\frac{\alpha_{2D_1}}{2}q)} \,,
\end{equation}
\begin{equation} \label{eq:2layer2}  
\Phi_{2D}(q,d)=\frac{Q/q \ee^{-dq}-\varepsilon_0\alpha_{2D_1}q e^{-dq}\Phi_{2D}(q,0)}{2 \varepsilon_0(1+\frac{\alpha_{2D_2}}{2}q)} \, .
\end{equation}
These expressions provide an intuitive understanding of the problem. On the one hand, a test charge in the lower layer is subject to the total charge $Q$ and experiences an additional induced charge density of the upper layer that exponentially decays with the interlayer distance $d$. On the other hand, a test charge in the upper layer senses both $Q$ and the induced charge density with an attenuation factor. 

We can now solve \cref{eq:2layer1} and \cref{eq:2layer2}.
Exemplary for $\Phi(q,0)$ this yields
\begin{equation} \label{bilayer 11}
    \begin{aligned}
        \Phi_{2D}(q,0)&=\frac{Q/q}{2\varepsilon_0(1+r_{01}q)}  - \frac{Q \varepsilon_0 \alpha_{2D_2} \ee^{-2dq}}{(2\varepsilon_0)^2(1+r_{01}q)(1+r_{02}q)}+ \frac{ \varepsilon_0^2 \alpha_{2D_1}\alpha_{2D_2}\ee^{-2dq}q^2}{(2\varepsilon_0)^2(1+r_{01}q)(1+r_{02}q)} \Phi(q,0)\\
        &=\frac{(2\varepsilon_0)^2 (1+r_{01}q)(1+r_{02}q)}{(2\varepsilon_0)^2(1+r_{01}q)(1+r_{02}q)-\varepsilon_0^2 \alpha_{2D_1}\alpha_{2D_2}\ee^{-2dq}q^2}\left(\frac{Q}{2\varepsilon_0q(1+r_{01}q)}-\frac{Q\varepsilon_0 \alpha_{2D_2}\ee^{-2dq}}{(2\varepsilon_0)^2(1+r_{01}q)(1+r_{02}q)}\right)\\
        &=\frac{(1+r_{02}q)Q \ee^{2dq}-r_{02}Qq}{2\varepsilon_0 q \ee^{2dq}(1+r_{01}q)(1+r_{02}q)-2\varepsilon_0r_{01}r_{02}q^3}\,,
    \end{aligned} 
\end{equation}
with the \textit{screening length} $r_0\equiv \alpha_{2D}/2$.
Analogously, we find
\begin{equation} \label{bilayer 12}
\Phi_{2D}(q,d)=\frac{ Q \ee^{dq}}{2\varepsilon_0 \ee^{2dq}q(1+qr_{01})(1+qr_{02})-2\varepsilon_0q^3r_{01}r_{02}}\,.
\end{equation}
The results behave as expected in both the limit of $d\xrightarrow{}0$

 \begin{equation}
   \lim_{d\xrightarrow{}0} \Phi_{2D}(q)=\frac{Q}{2 \varepsilon_0 q(1+(r_{01}+r_{02})q)}\,,
\end{equation}

 \noindent and $d\xrightarrow{}\infty$
\begin{equation}
    \lim_{d\xrightarrow{}\infty}\Phi_{2D}(q,0)=\frac{Q}{2 \varepsilon_0q(1+r_{01}q)}\,,
\end{equation}
which is exactly the monolayer potential \cite{Keldysh1979,Cudazzo2011}.
 
 The Fourier transformation to real space is performed numerically. Here, it is important to realize that in these results only the first and second component of momentum and position appear, while the third one only comes in as the concrete value of the layer separation $d$.

\section{Three-body Hamiltonian in the relative coordinate frame}

The bilayer Hamiltonian reads
\begin{equation}
\hat{H} = \sum_{i=1}^{3} \left[ \frac{-\hbar^2}{2 m_i} \Delta_{\vec{R}_i}-  \frac{Q_i \Delta E}{2} (\mathbb{1}-\tau_z^i) + t_i \tau_x^i \right] +\sum_{a,b}\sum_{i<j} V_{ab}(|\hat{\vec{R}}_i-\hat{\vec{R}}_j|) ~,
\end{equation}
with the mass $m_i$ and the spatial coordinate vector $\vec{R}_i$ of the $i$-th particle. The interaction potential between particles in layer $a$ and $b$ is given by $V_{ab}$.
Next we separate the center-of-mass motion via the transformation
\begin{equation}
\begin{split}
\vec{r}_1=&\vec{R}_1-\vec{R}_3~,~~~~~\vec{r}_2=\vec{R}_2-\vec{R}_3~,\\
\vec{r}_{\t{CM}}=&\frac{1}{M}(m_1\vec{R}_1+m_2\vec{R}_2+m_3\vec{R}_3)~,
\end{split}
\end{equation}
with total mass $M$, leading to the Hamiltonian
\begin{equation}
\begin{split}
\hat{H} = &-\sum_{i=1}^{2} \frac{\hbar^2}{2 \mu_i} \Delta_{\vec{r}_i} - \frac{\hbar^2}{m_3} \nabla_{\vec{r}_1}\cdot \nabla_{\vec{r}_2} -\frac{\hbar^2}{2M} \Delta_{r_{\t{com}}}
- \sum_{i=1}^{3} \frac{q_i \Delta E}{2} (\mathbb{1}-\tau_z^i) + t_i \tau_x^i
+ \sum_{a,b}\sum_{i=1}^{2} V_{ab}(|\hat{\vec{r}}_i|) + V_{ab}(|\hat{\vec{r}}_1-\hat{\vec{r}}_2|)~,
\end{split}
\end{equation}
where $\mu_i=m_im_3/(m_i+m_3)$ is the reduced mass.
Moreover, $r_1$, $r_2$ are the electron-hole distances while the angels $\alpha$ and $\theta$ determine the lab frame orientation and angle spanned by the majority charges, respectively \cite{Fey2020}.
Using these coordinates to parametrize $\vec{r}_1 = r_1(\t{cos}(\alpha +\theta/2)\t{\textbf{e}}_x +\t{sin}(\alpha + \theta/2)\t{\textbf{e}}_y)$ and $\vec{r}_2 = r_1(\t{cos}(\alpha -\theta/2)\t{\textbf{e}}_x +\t{sin}(\alpha - \theta/2)\t{\textbf{e}}_y)$ one arrives at the Hamiltonian in relative coordinates,
\begin{equation} 
\begin{split}
\hat{H}_{\t{rel}}^{\t{kin}}= & \sum_{i=1}^{2} \frac{-\hbar^2}{2 \mu_i} \left( \partial_{r_i}^2+\frac{\partial_{r_i}}{r_i}+\frac{\partial_{\theta}^2}{r_i^2} + \frac{\partial_{\alpha}^2}{4r_i^2}-\frac{\partial_{\theta}\partial_{\alpha}}{r_i^2} \right)\\& - \frac{\hbar^2}{m_3}\left[ \t{cos}\theta\, \partial_{r_1}\partial_{r_2} -\frac{\t{cos} \theta \, \partial_{\theta}^2}{r_1r_2} \right. \left. - \left( \frac{\partial_{r_1}}{r_2}+\frac{\partial_{r_2}}{r_1} \right) \t{sin}\theta\, \partial_{\theta} \right. \left. +\frac{\t{cos}\theta\, \partial_{\alpha}^2}{4r_1 r_2} +\left( \frac{\partial_{r_2}}{r_1}-\frac{\partial_{r_1}}{r_2} \right)\frac{\t{sin}\theta\, \partial_{\alpha}}{2}\right]
\end{split}
\end{equation}

The ansatz for the relative wave functions 
$$
\Psi(r_1,r_2,\theta,\alpha)=\psi(r_1,r_2,\theta) \ee^{\ii m \alpha}
$$
captures the conservation of angular momentum $m$.
In the following we focus on the zero angular momentum case $m=0$. 
The kinetic part of the Hamiltonian acting on the $m=0$ subspace is then given by
\begin{equation} 
\begin{split} 
\hat{H}_{\t{rel}}^{\t{kin}}= & \sum_{i=1}^{2} \frac{-\hbar^2}{2 \mu_i} \left( \partial_{r_i}^2+\frac{\partial_{r_i}}{r_i}+\frac{\partial_{\theta}^2}{r_i^2}\right)- \frac{\hbar^2}{m_3}\left( \t{cos}\theta\, \partial_{r_1}\partial_{r_2} -\frac{\t{cos} \theta \, \partial_{\theta}^2}{r_1r_2} \right.\left.
- \left( \frac{\partial_{r_1}}{r_2}+\frac{\partial_{r_2}}{r_1} \right) \t{sin}\theta\, \partial_{\theta} \right)~.
\end{split}
\end{equation}
The total Hamiltonian is then given by
\begin{equation}
\begin{split}
\hat{H} = & \hat{H}_{\t{rel}}^{\t{kin}}- \sum_{i=1}^{3} \frac{q_i \Delta E}{2} (\mathbb{1}-\tau_z^i) + t_i \tau_x^i+ \sum_{a,b}\sum_{i=1}^{2} V_{ab}(r_i) + V_{ab}\left(\sqrt{r_1^2+r_2^2-2r_1r_2\,\t{cos}\,\theta}\right)~.
\end{split}
\end{equation}

\section{Discrete Variable Representation (DVR)}

In order to numerically calculate eingenvalues and eigenstates of a given Hamiltonian a concrete choice of basis functions is needed for the matrix representation of the Hamiltonian. 
Here we present details on the choice of the basis functions used in the main text to represent the three-body Hamiltonian numerically allowing for its diagonalization.
Thereby we summarize and follow the discussion given in \cite{Beck2000} and \cite{Fey2020}.

First we consider an infinite set of square-integrable basis functions $\{ \varphi_j(x) \}_{j=1}^{\infty}$. Each pair $\varphi_j$ and $\varphi_k$ allows to analytically evaluate the matrix elements
\begin{equation} \label{eq:MatElements}
\begin{split}
Q_{jk} & = \bra{\varphi_j} \hat{x} \ket{\varphi_k} ~,\\
D_{jk}^{(1)} & = \bra{\varphi_j} \frac{\dd}{\dd x} \ket{\varphi_k}~,\\
D_{jk}^{(2)} & = \bra{\varphi_j} \frac{\dd^2}{\dd x^2} \ket{\varphi_k}~,
\end{split}
\end{equation}
that usually appear in a typical Hamiltonian.
Here it is convenient to also assume square-integrability of $x \, \varphi_j(x)$ and $\dd/\dd x \, \varphi_j(x)$.

In the following we will investigate the truncated basis $\{ \varphi_j(x) \}_{j=1}^{N}$ formed by the first $N$ functions. 
The projector
\begin{equation}
	\hat{P} = \sum_{j=1}^{N} \ket{\varphi_j}\bra{\varphi_j}
\end{equation}
projects any state on the subspace spanned by the truncated basis.
Further we define the matrices\footnote{Note, in the following matrices will be labeled by bold capital latin letters.} $\vec{Q}$, $\vec{D}^{(1)}$ and $\vec{D}^{(2)}$ acting on this subspace by allowing only $j,k \leq N$ in \cref{eq:MatElements}. 
 
If the potential energy is the expectation value of an operator that itself is a function of the position operator $\hat{x}$, i.e.,
\begin{equation}
	\hat{V}=V(\hat{x})~,
\end{equation}
there are two different ways of approximately representing $\hat{V}$.  
First, the matrix can be defined in the truncated basis via the matrix elements
\begin{equation} 
	V_{jk}^{\t{VBR}} = \bra{\varphi_j}\hat{V}\ket{\varphi_k}~,
\end{equation}
with $j,k \leq N$ or as a function of the matrix representing the position operator in the truncated basis, i.e.,
\begin{equation}
\vec{V}^{\t{FBR}} = V(\vec{Q})~.
\end{equation}
Here VBR refers to `variational basis-set representation' and FBR refers to `finite basis-set representation'.
Note that $\vec{V}^{\t{FBR}}$ may have components that are outside of the subspace spanned by the truncated basis and that $\vec{V}^{\t{VBR}}=\vec{V}^{\t{FBR}}$ in general only holds in the limit $N \rightarrow \infty$, where both representations are exact.

The Hamiltonian 
\begin{equation}
	\hat{H} = \hat{T} + \hat{V}
\end{equation}
can be approximated using either of the approximate representations and the resulting bound state energies represent upper bounds to the exact (infinite dimensional) problem.

In general the evaluation of the matrix elements $V_{jk}^{\t{VBR}}$ is complicated, whereas the $\vec{V}^{\t{FBR}}$ is easier to handle as $\vec{Q}$ is diagonalizable
\begin{equation}
	\vec{Q}=\vec{U}\, \vec{X} \, \vec{U}^{\dagger}~.
\end{equation} 
Here $\vec{X}=\t{diag}(x_1,x_2,...x_N)$ is the diagonalized matrix, where $x_{\alpha}$ are the eigenvalues of $\vec{Q}$ and $\vec{U}$ is the matrix containing the respective eigenvectors $\ket{u_{\alpha}}$.
Consequently, the matrix $\vec{V}^{\t{FBR}}$ is obtained as
\begin{equation}
	\vec{V}^{\t{FBR}} = V(\vec{Q}) = \vec{U}\, V\!(\!\vec{X}\!\hspace*{0.3mm}) \, \vec{U}^{\dagger} =  \vec{U}\, \t{diag}\big(V(x_1),V(x_2),...,V(x_N)\big) \, \vec{U}^{\dagger}~,
\end{equation}
or element-wise 
\begin{equation}
	V_{jk}^{\t{FBR}}= 
	\sum_{\alpha=1}^{N} U_{j \alpha} V(x_{\alpha}) U_{k \alpha}^{\ast}~.
\end{equation}

The \textit{discrete variable representation} (DVR) basis is now the basis in which the potential energy operator's matrix representation is diagonal, i.e.,
\begin{equation}
\begin{split}
	\vec{V}^{\t{DVR}} = & \vec{U}^{\dagger} \, \vec{V}^{\t{FBR}} \, \vec{U} = \vec{U}^{\dagger} \vec{U}\, V\!(\!\vec{X}\!\hspace*{0.3mm}) \, \vec{U}^{\dagger} \vec{U} = V(\vec{X})\\
	 = & \t{diag}\big(V(x_1),V(x_2),...,V(x_N)\big) ~.
\end{split}
\end{equation}
Here the eigenvalues $x_{\alpha}$ of the position operator $\vec{Q}$ appear as lattice sites on which the potential is evaluated.
The corresponding basis is spanned by superpositions of the truncated basis functions, i.e., $\{ \varphi_j(x) \}_{j=1}^{N}$, where the weights are given by the matrix elements of the unitary transformation $\vec{U}$
\begin{equation}
	\ket{\chi_{\alpha}(x)} = 
		\sum_{\alpha=1}^{N} \varphi_j(x) U_{j \alpha}~.
\end{equation}

To obtain the numerical results in the main text we use 11 Laguerre functions $\varphi_j$ in each radial direction and seven in the angular direction.

\section{Layer distance dependence}

\begin{figure}
        \centering
		\includegraphics{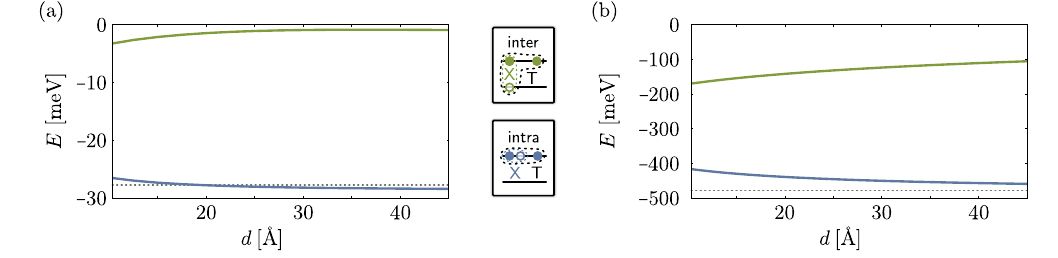}
		\caption{
Layer distance dependence of (a) the trion and (b) the exciton binding energies (interlayer in green, intralayer in blue) starting at the value of $d=10.3$\,\AA\ corresponding to the presence of a single hBN layer in between the two semiconductors. The results for the decoupled interlayer and intralayer configurations are obtained for a vanishing hole tunnelling amplitude, i.e.\ $t_3=t=0$.
(a) For large values of $d$ the interlayer trion binding energy approaches zero. 
The intralayer trion binding energy, in contrast, approaches the value of the monolayer case $-27.7$ meV (horizontal line) found in Ref.~\cite{Fey2020}. The small offset is caused by the use of a smaller set of DVR basis functions, which was 
introduced due to the increased size of Hilbert space in the multilayer problem compared to the monolayer investigated in \cite{Fey2020}.
(b) For large $d$ the intralayer exciton binding energy approaches the value of the monolayer case $476.7$ meV (horizontal line) found in Ref.~\cite{Fey2020}. 
 }
		\label{fig:BindingEnergies}
	\end{figure}

Here we numerically investigate the influence of layer separation on exciton and trion binding energies and check for the convergence of the bilayer results to the monolayer scenario in the limit of large $d$. 
In general, multilayer binding energies differ from the monolayer scenario due to the additional dielectric screening caused by the presence of the second TMD layer.
Fig.\ \ref{fig:BindingEnergies} shows the bilayer trion (a) and exciton binding energies (b).

We find that the interlayer trion remains bound in a significant window of $d$ values, including the case of a single hBN monolayer studied in the main text. 
The mechanism of two-dimensional exciton-electron Feshbach resonances presented in the main text works as long as the respective trions exist (and the band structure offset between the TMD layers is tunable with respect to the other layer via, e.g., an external electric field). 
This shows that the exact value of $d$ is not crucial for observing Feshbach resonances. 

As the distance of the two TMD layers becomes large, and the additional dielectric screening effect of the second layer becomes small, both the intralayer exciton and trion binding energies approach the monolayer results found in \cite{Fey2020} which serves as further benchmark of our method (remaining deviations arise due to the different number of the basis functions we have used in the numerics). 
The interlayer binding energies in contrast vanish in the limit of large $d$ as the respective charges are increasingly separated.